\documentclass[draft]{agujournal2019}
\usepackage{url} 
\usepackage{lineno}
\graphicspath{{../figures/}}
\journalname{JGR: Planets}

\begin{document}

%
%


\title{The Dayside Ionopause of Mars: Solar Wind Interaction, Pressure Balance, and Comparisons with Venus}

%
%




\authors{F. Chu\affil{1}, Z. Girazian\affil{2}, F. Duru\affil{3}, R. Ramstad\affil{4}, J. Halekas\affil{2}, D. A. Gurnett\affil{2}, X. Cao\affil{4}, and A. J. Kopf\affil{5}}


\affiliation{1}{Physics Division, Los Alamos National Laboratory, Los Alamos, NM, USA}
\affiliation{2}{Department of Physics and Astronomy, University of Iowa, Iowa City, IA, USA}
\affiliation{3}{Department of Physics, Coe College, Cedar Rapids, IA, USA}
\affiliation{4}{Laboratory for Atmospheric and Space Physics, University of Colorado Boulder, Boulder, CO, USA}
\affiliation{5}{U.S. Naval Observatory, Washington, DC, USA}





\correspondingauthor{F. Chu}{fchu@lanl.gov}




\begin{keypoints}
\item The ionopauses at Mars are found 13\% of the time unmagnetized; this percentage at Venus, however, is up to 65\%
\item The ionopause altitude decreases with the solar wind dynamic pressure at Mars, similar to the altitude variation of the ionopauses at Venus
\item The ionopause thickness at Mars and Venus is mainly determined by the ion gyromotion and equivalent to about 5 ion gyroradii
\end{keypoints}




%
%


\begin{abstract}
Due to the lower ionospheric thermal pressure and existence of the crustal magnetism at Mars, the Martian ionopause is expected to behave differently from the ionopause at Venus. We study the solar wind interaction and pressure balance at the ionopause of Mars using both in situ and remote sounding measurements from the MARSIS (Mars Advanced Radar for Subsurface and Ionosphere Sounding) instrument on the Mars Express orbiter. We show that the magnetic pressure usually dominates the thermal pressure to hold off the solar wind at the ionopause at Mars, with only 13\% of the cases where the ionospheric thermal pressure plays a more important role in pressure balance. This percentage at Venus, however, is up to 65\%. We also find that the ionopause altitude at Mars decreases as the normal component of the solar wind dynamic pressure increases, similar to the altitude variation of the ionopauses at Venus. Moreover, our results show that the ionopause thickness at Mars and Venus is mainly determined by the ion gyromotion and is equivalent to about 5 ion gyroradii.
\end{abstract}

\section*{Plain Language Summary}
An ionopause is a sharp decrease in the plasma density at the top of the ionosphere, separating the ionospheric plasma from the shocked solar wind plasma. It was found to be a common feature at Venus, where its variability is well constrained by observations from the Pioneer Venus Orbiter. Past studies have shown that there are many similarities between the ionopauses at Mars and Venus. However, because the thermal pressure of the ionosphere at Mars is lower than that of Venus, the Martian ionopause is also thought to behave differently from the ionopause at Venus. We study the pressure configuration inside the ionopause at Mars and find that most of the time the magnetic pressure is greater than the thermal pressure. We also find that higher solar wind pressure pushes the ionopause downward at Mars. Moreover, we show that the thickness of the ionopauses at Mars equals a few radii of ion circular motion in the magnetic field. Our results provide insight to the process that controls the formation of the ionopause at Mars.

%
%

%


%
%
%
%

\section{Introduction}
\label{sec:intro}

Since Mars does not possess a strong global intrinsic magnetic field, the incident solar wind plasma and interplanetary magnetic field interacts inductively with its upper atmosphere and highly conductive ionosphere. This interaction induces current sheets that produce a magnetic barrier to prevent the solar wind from further penetrating into the atmosphere \cite{ramstad_global_2020}, resulting in the formation of several plasma boundaries around Mars, such as the magnetic pileup boundary \cite{crider_observations_2002, bertucci_mgs_2004, bertucci_structure_2005}, photoelectron boundary \cite{garnier_martian_2017, han_relationship_2019}, and the ionopause \cite{nagy_plasma_2004, han_discrepancy_2014}.

The Martian ionopause is a feature defined as a steep gradient in electron density at the topside of the ionosphere. It is a tangential discontinuity that marks the transition from the hot plasma in the induced magnetosphere to the cold, dense ionospheric plasma. When the solar wind flows around Mars, it exerts its dynamic pressure indirectly on the ionosphere through the magnetosheath and magnetic pileup region. Therefore, as the ionospheric thermal pressure decreases sharply across the ionopause, a magnetic pressure from an intrinsic or induced field is required to maintain a steady state of the ionosphere \cite{holmberg_maven_2019, sanchezcano_mars_2020}.


Since Venus also lacks a global-scale magnetic field and its upper atmosphere interacts directly with the solar wind, the ionopauses at Venus and Mars are similar in many aspects. For example, solar wind conditions can heavily influence the dynamics of the ionopauses at both planets \cite{phillips_dependence_1985, sanchezcano_mars_2020}. Moreover, the altitudes of Venusian and Martian ionopauses both vary over the solar cycle as a result of the periodic effects of solar EUV on the ionospheric thermal pressure \cite{kliore_solar_1991, duru_martian_2020}. Thanks to the Pioneer Venus Orbiter (PVO) mission, the studies of the ionopause at Venus have provided many deep insights that can be applied to understanding the formation of the ionopause at Mars.

On the other hand, due to the fact that Mars has a low ionospheric thermal pressure and strong localized crustal magnetism, the Martian ionopause is also strikingly different from that at Venus. For instance, unlike Venus, the ionosphere at Mars is usually found to be in a magnetized state because the solar wind dynamic pressure often exceeds the maximum ionospheric thermal pressure \cite{nagy_plasma_2004}. The magnetic pressure, therefore, plays a more important role in standing off the solar wind ram pressure at the ionopause at Mars. In addition, fewer ionopauses are found over strong crustal magnetic field regions at Mars as minimagnetosphere can often form to prevent the solar wind from reaching the ionosphere. For example, \citeA{chu_effects_2019} showed that most of the ionopauses are detected in locations where the crustal field strength at 400 km is less than 40 nT. \citeA{sanchezcano_mars_2020} also found that the ionopauses occur at 36\% of the time over strong crustal magnetic fields compared with 54\% of the time over regions with weak magnetic fields.

In the past, a number of studies have been dedicated to the investigation of the dependence of the ionopause altitude on solar zenith angle (SZA) and solar extreme ultraviolet (EUV) flux, as well as the effects of the crustal magnetic fields on the formation of the ionopause at Mars \cite{vogt_ionopause-like_2015, chu_effects_2019, duru_martian_2020}. In this paper, we take advantage of both in situ and remote sounding measurements from the Mars Advanced Radar for Subsurface and Ionosphere Sounding (MARSIS) instrument on board the Mars Express (MEX) spacecraft to study the pressure configuration and balance in the Martian ionopause. We also report for the first time on the mechanisms that control the thickness of the ionopause at Mars. The paper is organized as follows: section~\ref{sec:ionopause} gives a description of the ionopause observations from the MARSIS instrument, section~\ref{sec:pressure} explains the pressure terms used in the analysis, section~\ref{sec:results} presents the results, section~\ref{sec:discussion} gives a discussion on ionopause thickness, and section~\ref{sec:conclusions} concludes the paper.

\section{Ionopause Observations}
\label{sec:ionopause}

The MARSIS instrument on board the Mars Express spacecraft is a low-frequency radar sounder designed to perform both subsurface and ionospheric soundings \cite{picardi_marsis_2004}. In this study, the ionopause is detected by the Active Ionospheric Sounding mode of the MARSIS radar using two different techniques -- topside remote radar sounding and in situ measurements.


Spacecraft-borne ionospheric sounders take advantage of the fact that electromagnetic waves will be reflected at the surface where the wave frequency is below the local electron plasma frequency
\begin{equation}
\label{eq:fpe}
f_\textup{\scriptsize{pe}}=8980\sqrt{n_\textup{\scriptsize{e}}}\textup{ Hz},
\end{equation}
where $n_\textup{\scriptsize{e}}$ is the electron density in $\textup{cm}^{-3}$ \cite{hagg_interpretation_1969}. The MARSIS remote radar sounding is performed by transmitting a short radio pulse towards the ionosphere and detecting the time delay, $\Delta t$, for the echo reflected off the ionosphere to return. The apparent altitude or virtual height \cite{wright_automatic_1972} of the reflection point can be expressed as
\begin{equation}
\label{eq:h}
h=h_\textup{\scriptsize{MEX}}-\frac{c\Delta t}{2},
\end{equation}
where $h_\textup{\scriptsize{MEX}}$ is the spacecraft altitude and $c$ is the speed of light. Therefore, by stepping through the frequency of the transmitting radar pulses (0.1 to 5.5 MHz), MARSIS is able to obtain the vertical plasma density profile of the ionosphere from the spacecraft location to the peak of the electron density \cite{gurnett_radar_2005}. The term ``apparent altitude'' here refers to the scale that has not been corrected for dispersion of the electromagnetic waves propagating in a plasma. Since the discrepancy between the actual ionopause altitude and the apparent altitude becomes larger as MEX is further away from the ionopause during remote sounding, here we only consider the cases where the spacecraft altitude is below 700 km. Choosing such a subdataset will not change our results much as 96\% of the ionopauses occur in the altitude range between 300 and 430 km \cite{chu_effects_2019}.

The intensities of the return echoes received by MARSIS during each frequency sweep and the time delay $\Delta t$ are displayed in ionograms. A schematic illustration of an electron plasma frequency profile as a function of altitude is shown in Figure~\ref{fig:ionogram}a and the resulting ionospheric sounding ionogram is shown in Figure~\ref{fig:ionogram}b. One of the most prominent features that can be easily seen in an ionogram is the ionospheric echo, usually extending from $\sim$ 1 MHz to a few MHz. An ionopause echo, on the other hand, appears much less often in the ionogram. In Figure~\ref{fig:ionogram}b, an ionopause is identified as a short horizontal line in the low frequency range ($<$ 0.4 MHz), indicating a steep density change over a short vertical distance. A potential problem we initially thought of in identifying the ionopause in an ionogram is that the small ionopause signature could sometimes be obscured by vertical or horizontal lines in the same region (electron plasma frequency harmonics or electron cyclotron echoes). After a careful examination, however, we find that this is not an issue since these lines do not appear in the ionogram anymore when the local plasma density at the spacecraft location is well below 100 $\textup{cm}^{-3}$, a condition satisfied most of the time during MARSIS remote sounding from above the ionopause. More detailed discussions about identifying the ionopause in an ionogram can be found in \citeA{chu_effects_2019}.

Besides the remote sounding technique, the ionopause can also be detected through MARSIS in situ density measurements \cite{duru_steep_2009}. In the process of remote sounding, intense electrostatic electron plasma oscillations can be excited at the local electron plasma frequency surrounding the spacecraft \cite{gurnett_overview_2008}. Later when these oscillations are picked up by MARSIS, the received waveforms are often severely clipped, resulting in closely spaced vertical harmonic lines in the low frequency region of the ionogram (Figure~\ref{fig:ionogram}c). These harmonics allow us to determine the local electron density
\begin{equation}
\label{eq:h}
n_\textup{\scriptsize{e}}=\left ( \frac{\Delta f}{8980} \right )^2\textup{ cm}^{-3},
\end{equation}
where $\Delta f$ is the harmonic spacing in the units of Hz. As MEX enters or exits the topside ionosphere, one can identify the ionopause by looking for the signature of a steep density gradient in the local electron density profile.



Both of these two techniques, remote radar sounding and in situ measurements, have advantages and disadvantages in ionopause observations. MARSIS remote sounding can detect the ionopause continuously from above the ionosphere, whereas the in situ method can only observe the ionopause twice during each orbit. In our remote sounding dataset, $\sim$ 1,600 out of 1,791 ionopauses are observed in completely different orbits, accounting for $\sim$ 89\% of all the detections. On the other hand, the remote sounding technique cannot measure the magnetic field inside the ionopause. The in situ method therefore becomes the only way for us to investigate the pressure configuration at the ionopause. Despite the advantages of these two techniques, the ionopause is still a transient feature of the Martian system and only observed using this technique less than 18\% of the time \cite{duru_steep_2009, chu_effects_2019}. It should also be noted that neither of these two techniques can measure the vertical plasma density profile in ``pure'' vertical direction due to the 3D nature of the spacecraft orbit. 



\section{Pressure Terms at the Ionopause}
\label{sec:pressure}

To investigate the solar wind interaction and pressure balance at the ionopause, we need to evaluate the pressures that are exerted on the ionopause. In this study specifically, we consider three different pressure terms, the thermal pressure and magnetic pressure in the ionosphere, and the solar wind dynamic pressure.


\subsection{Ionospheric Thermal Pressure}
\label{subsec:itp}

\begin{figure*}
\begin{center}
\includegraphics[width=6.69in]{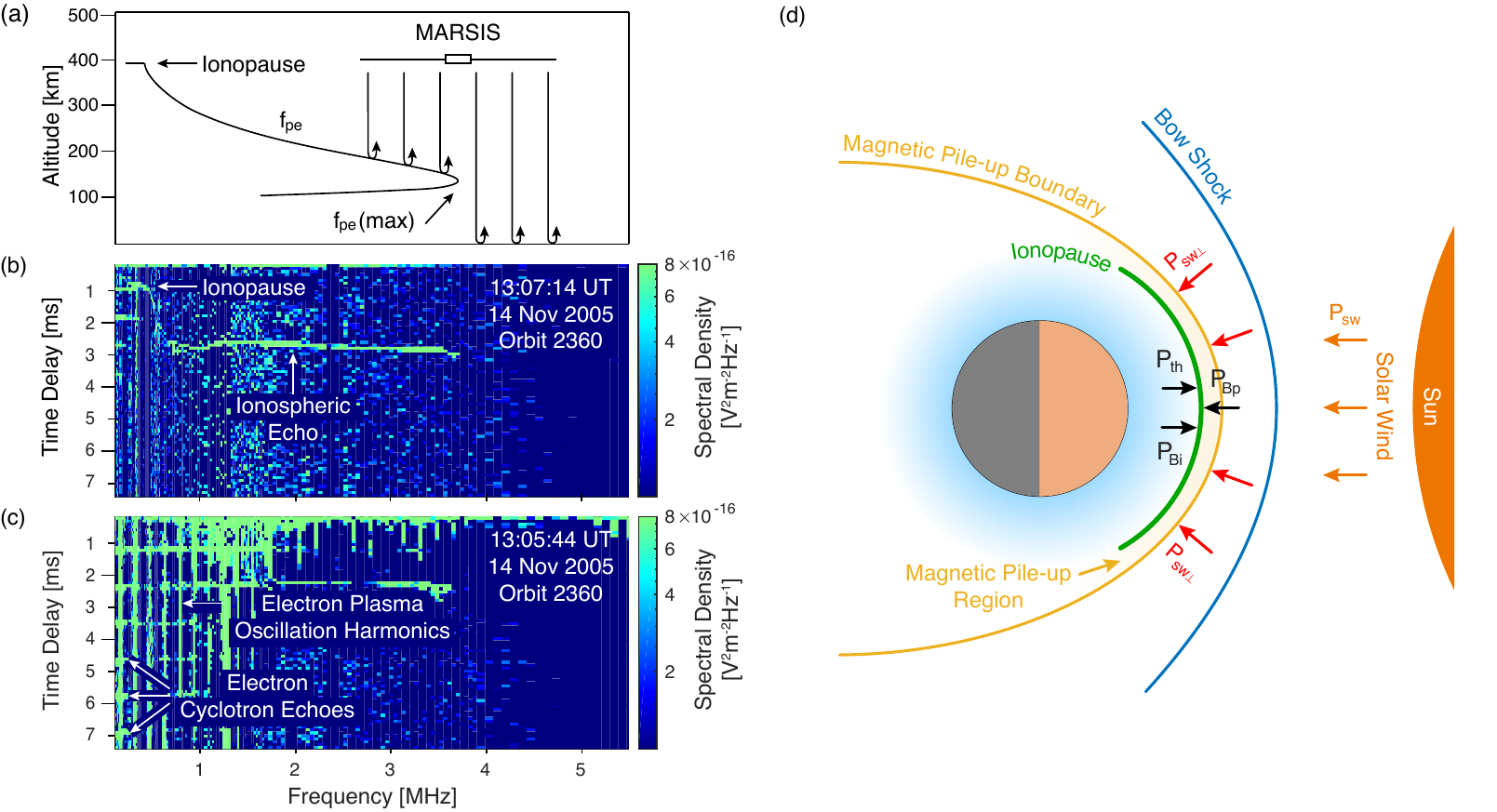}
\caption{(a) Schematic illustration of a typical electron plasma frequency profile as a function of altitude in the Martian ionosphere. (b) An example of the color-coded MARSIS ionogram from the orbit 2360 on 14 November 2005. The ionopause can be seen as a horizontal line at frequencies below 0.4 MHz. (c) Another ionogram from the same orbit showing the features of the electron plasma oscillation harmonics and electron cyclotron echoes. (d) Schematic illustration of the pressure terms at the ionopause. $P_\textup{\scriptsize{th}}$ is the ionospheric thermal pressure, $P_\textup{\scriptsize{Bi}}$ is the magnetic pressure in the ionosphere, $P_\textup{\scriptsize{Bp}}$ is the magnetic pressure in the magnetic pileup region, and $P_\textup{\scriptsize{sw}}$ is the solar wind dynamic pressure.}
\label{fig:ionogram}
\end{center}
\end{figure*}


Based on MARSIS in situ electron density measurements, the thermal pressure of the ionosphere can be estimated as
\begin{equation}
\label{eq:pth}
P_\textup{\scriptsize{th}}=n_\textup{\scriptsize{i}}kT_\textup{\scriptsize{i}}+n_\textup{\scriptsize{e}}kT_\textup{\scriptsize{e}} \approx 2n_\textup{\scriptsize{e}}kT_\textup{\scriptsize{e}},
\end{equation}
where $T_\textup{\scriptsize{i}}$ and $T_\textup{\scriptsize{e}}$ are the ion and electron temperature, respectively, $n_\textup{\scriptsize{i}}$ is the ion density ($n_\textup{\scriptsize{i}} \approx n_\textup{\scriptsize{e}}$), and $k$ is the Boltzmann constant. Here we assume equal ion and electron temperature $T_\textup{\scriptsize{i}} \approx T_\textup{\scriptsize{e}}$ as the first order approximation for at least up to 350 km \cite{hanson_viking_1988, matta_numerical_2014}. We also use a fixed representation of the $T_\textup{\scriptsize{e}}$ profile that does not account for potential SZA, latitude, seasonal, or solar activity variations \cite{ergun_dayside_2015, pilinski_electron_2019}: 
\begin{equation}
\label{eq:te}
T_\textup{\scriptsize{e}}=\frac{T_\textup{\scriptsize{H}}+T_\textup{\scriptsize{L}}}{2}+\frac{T_\textup{\scriptsize{H}}-T_\textup{\scriptsize{L}}}{2}\tanh\left ( \frac{z-Z_0}{H_0} \right ),
\end{equation}
where $T_\textup{\scriptsize{H}}=0.271$ eV and $T_\textup{\scriptsize{L}}=0.044$ eV are the upper and lower bounds of $T_\textup{\scriptsize{e}}$, respectively, $z$ is the altitude, $Z_0=241$ km represents the altitude of the most rapid change in $T_\textup{\scriptsize{e}}$, and $H_0=60$ km is the scale height of the rapid change. 


\subsection{Magnetic Pressure in the Ionosphere}
\label{subsec:magnetic}

In addition to the vertical electron plasma oscillation harmonics discussed in section~\ref{sec:ionopause}, another commonly found feature in many MARSIS ionograms is a series of equally spaced horizontal echoes in time at frequencies below 1 MHz, as shown in Figure~\ref{fig:ionogram}c. When electrons near the antenna are accelerated by the strong electric fields during each transmission cycle, they go through cyclotron motions in the local magnetic field and periodically return to the vicinity of the antenna, causing the electron cyclotron echoes to appear in the ionogram \cite{gurnett_radar_2005}. The repetition rate of these echoes is equal to the local electron cyclotron frequency
\begin{equation}
\label{eq:B}
f_\textup{\scriptsize{c}}=\frac{Be}{2\pi m_\textup{\scriptsize{e}}},
\end{equation}
where $B$ is the magnetic field strength, $e$ is the electron charge, and $m_\textup{\scriptsize{e}}$ is the electron mass. Since MEX is not equipped with a magnetometer, these electron cyclotron echoes provide the only method to measure the local magnetic field. All the ionopause in situ detections selected in this study are accompanied by simultaneous magnetic field measurements.

As the peak thermal pressure in the Mars ionosphere rarely exceeds the solar wind dynamic pressure, the dayside ionosphere is often found to be magnetized in order to stand off the solar wind \cite{zhang_post-pioneer_1990, nagy_plasma_2004}. Over the regions away from the strong crustal magnetism, we can assume that the magnetic field is approximately tangential to the ionopause \cite{halekas_flows_2017}. Thus, the magnetic pressure in the ionosphere normal to the ionopause can be estimated to the first order as
\begin{equation}
\label{eq:pb}
P_\textup{\scriptsize{Bi}}=\frac{B^2}{2\mu _0},
\end{equation}
where $\mu _0$ is the permeability of free space.

\subsection{Solar Wind Dynamic Pressure}
\label{subsec:swdp}

Since MEX does not directly measure the properties of the solar wind, the solar wind dynamic pressure $P_\textup{\scriptsize{sw}}$ is estimated based on the ASPERA (Analyzer of Space Plasma and Energetic Atoms) solar wind moments, which are calculated from averaged proton distributions collected over the inbound/outbound segments of MEX outside the bow shock \cite{barabash_analyzer_2006, ramstad_martian_2015, ramstad_solar_2017}
\begin{equation}
\label{eq:psw}
P_\textup{\scriptsize{sw}}=n_\textup{\scriptsize{p}}m_\textup{\scriptsize{p}}v_\textup{\scriptsize{p}}^2,
\end{equation}
where $n_\textup{\scriptsize{p}}$ is the proton density, $m_\textup{\scriptsize{p}}$ is the proton mass, and $v_\textup{\scriptsize{p}}$ is the speed of the solar wind. As ASPERA makes measurements in the solar wind just before entering the bow shock and after exiting it, due to the MEX orbital period, the pressure values used in this study are the closest ones to the time when the ionopause is observed within 6 hours. 

Note that the normal component of the solar wind ram pressure is not directly exerted on the ionopause; rather it is first converted to thermal pressure in the magnetosheath, then to magnetic pressure in the pileup region. The normal component of the solar wind dynamic pressure can be written as
\begin{equation}
\label{eq:psw}
P_{\textup{\scriptsize{sw}} \perp}=\alpha P_\textup{\scriptsize{sw}}\cos^2\theta,
\end{equation}
where $\theta$ is the angle between the magnetic pileup boundary normal and the flow direction of the upstream solar wind, and $\alpha \approx 0.88$ is the proportionality constant \cite{crider_proxy_2003}. For lower solar zenith angles, $\theta$ can be approximately replaced by SZA. At higher SZAs, however, this approximation breaks down because the curvature of the obstacle must be accounted. Therefore, we only select the dataset with $\textup{SZA}$ less than $65^\circ$ in this study. 

A schematic illustration of the three pressure terms $P_\textup{\scriptsize{th}}$, $P_\textup{\scriptsize{Bi}}$, and $P_\textup{\scriptsize{sw}}$ is shown in Figure~\ref{fig:ionogram}d. In theory, a pressure balance across the ionopause requires
\begin{equation}
\label{eq:pb}
P_\textup{\scriptsize{th}}+P_\textup{\scriptsize{Bi}}=P_{\textup{\scriptsize{sw}} \perp}.
\end{equation}
In other words, the ionospheric thermal pressure and magnetic pressure should stand off the normal component of the solar wind dynamic pressure. A detailed examination of this pressure balance relation will be presented in section~\ref{sec:results}.

\section{Results}
\label{sec:results}

In this study, we utilize both MARSIS in situ (79 detections) and remote sounding (1,791 detections) measurements, excluding crustal magnetic field regions, to study the pressure balance at Martian ionopauses and their interactions with the solar wind. The descriptions of these datasets can be found in \citeA{chu_effects_2019} and \citeA{duru_martian_2020}. We first compare the pressure configuration at the ionopause at Venus and Mars, and then investigate the role of the solar wind in the formation of the Martian ionopause. Finally, we study the dependence of the ionopause thickness at Mars on altitude and magnetic field strength. The MARSIS remote sounding data are only used in Figure~\ref{fig:sw}a in section~\ref{subsec:influence} while the in situ data are used in the rest of the plots in this paper.

\subsection{Comparison of Pressure Configuration at Ionopauses at Venus and Mars}
\label{subsec:com}

\begin{figure*}
\begin{center}
\includegraphics[width=6.69in]{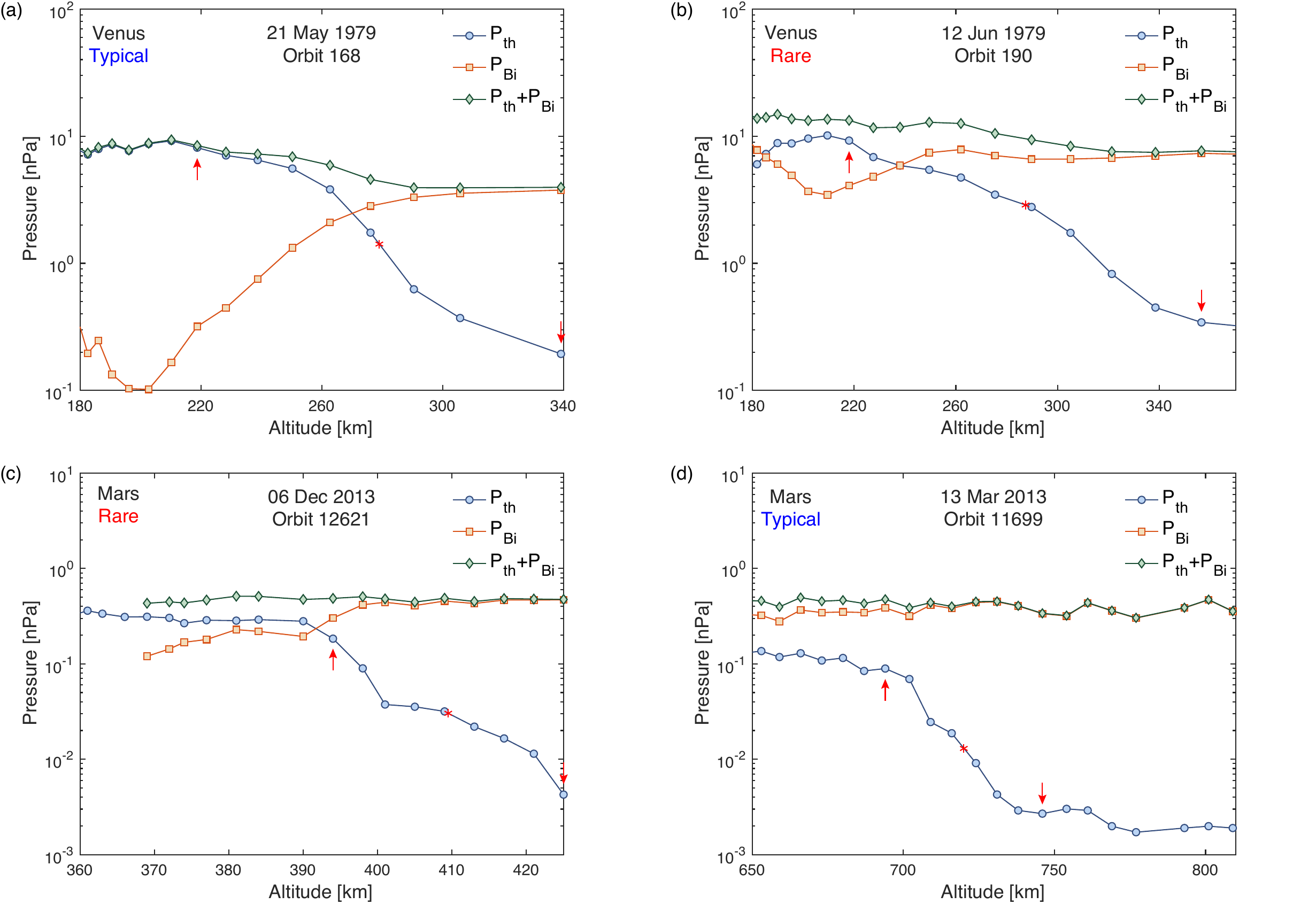}
\caption{(a)--(b) Thermal pressure and magnetic pressure profiles at the Venusian ionopause as a function of the altitude. The data shown here are based on the in situ measurements from the Pioneer Venus Orbiter's Magnetometer (OMAG) and Electron Temperature Probe (OETP). A typical example of the pressure configuration is shown in (a) and a rare case is shown in (b). (c)--(d) Thermal pressure and magnetic pressure profiles at the Martian ionopause as a function of the altitude. The data shown are based on the MARSIS in situ measurements. A rare, Venusian-like example of the pressure configuration is shown in (c) and a typical case is shown in (d). The red arrows mark the ionopause thickness and the red stars represent the location where the total pressure $P_\textup{\scriptsize{tot}}=P_\textup{\scriptsize{th}}+P_\textup{\scriptsize{Bi}}$ is measured.}
\label{fig:alt}
\end{center}
\end{figure*}


The maximum ionospheric thermal pressure at Venus often exceeds the solar wind dynamic pressure \cite{zhang_comparisons_1992}. The thermal pressure of the ionosphere, therefore, plays a dominant role in the pressure balance underneath the ionopause. Figure~\ref{fig:alt}a shows a typical example of the pressure configuration at the ionopause at Venus (Orbit 168 on 21 May 1979) during solar maximum. Inside the ionopause, as the thermal pressure decreases, the magnetic pressure is forced to increase to maintain the pressure balance. There are some rare occasions  shown in Figure 2b (Orbit 190 on 12 June 1979), where the magnetic pressure underneath the ionopause is around the same order of magnitude as the thermal pressure. In such cases, both of the pressure terms are equally important in balancing the solar wind dynamic pressure above the ionopause.


Due to the lower ionospheric thermal pressure at Mars, the behavior of the Martian ionopause is expected to be different from the ionopause at Venus. Figure 2c shows a rare, Venusian-like (typical pressure configuration at Venus; unmagnetized) pressure configuration at the Martian ionopause, only found in 13\% of all the cases over a 11-year period. This type of configuration usually occurs when the normal component of the solar wind dynamic pressure is extremely low ($<0.03$ nPa), so that the ionosphere is mostly in an unmagnetized state as the maximum thermal pressure of the ionosphere is sufficient to balance the total pressure above the ionopause, much like the case at Venus. In contrast, the similar pressure configuration at Venus shown in Figure~\ref{fig:alt}a is observed at least 65\% of cases at solar maximum \cite{luhmann_observations_1980, elphic_venus_1981} and $\sim 52\%$ at solar minimum \cite{angsmann_magnetic_2011}. Figure~\ref{fig:alt}d shows a typical pressure configuration (magnetized) at the ionopause at Mars, where the magnetic pressure is clearly seen to play a dominant roll in the pressure balance anywhere at the ionopause.

\subsection{Influence of Solar Wind on Martian Ionopause}
\label{subsec:influence}

\begin{figure*}
\begin{center}
\includegraphics[width=6.69in]{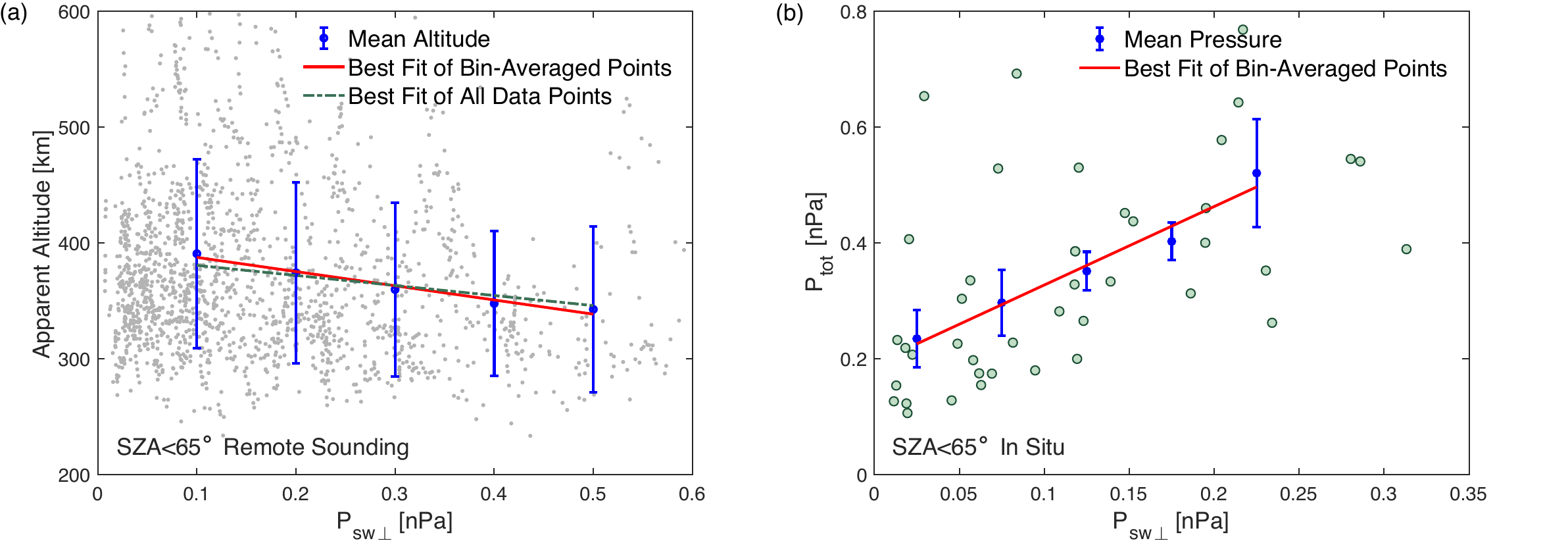}
\caption{(a) Scatter plot of the ionopause apparent altitude ($\textup{SZA}<65^\circ$) as a function of the normal component of the solar wind dynamic pressure at Mars. The ionopause data are collected using the MARSIS remote sounding technique. The mean ionopause altitude averaged over each 0.1 nPa bin is shown in blue dots. Error bars represent the standard deviation. (b) Correlation between the total ionopause pressure ($\textup{SZA}<65^\circ$) and normal component of the solar wind dynamic pressure at Mars. The mean ionopause pressure for each 0.05 nPa bin is shown in blue dots. Error bars represent the standard error of the mean ($\sigma / \sqrt{N}$, $\sigma$ standard deviation and $N$ number of data points in each bin). The best fit lines in (a) and (b) are based on the least squares method. SZA=solar zenith angle.}
\label{fig:sw}
\end{center}
\end{figure*}

At Venus, the behavior of the ionopause is strongly affected by the upstream solar wind conditions \cite{brace_structure_1991}. In our previous study, we showed how crustal magnetism and solar extreme ultraviolet flux control the ionopause formation at Mars \cite{chu_effects_2019}. Here, we examine the influence of the solar wind dynamic pressure on ionopause apparent altitude at Mars based on 1,791 ionopause detections ($\textup{SZA}<65^\circ$) obtained using the MARSIS remote sounding technique (Figure~\ref{fig:sw}a). Despite a strong scattering of the data points in Figure~\ref{fig:sw}a, likely due to the variations of seasons and solar extreme ultraviolet (EUV) flux (more plots showing the distribution in season, EUV flux, latitude, and SZA can be found in the supporting information), we find that the ionopause altitude decreases with the normal component of the solar wind dynamic pressure by $131 \pm 50$ km/nPa based on the least squares fitting of the bin-averaged points with 95\% confidence intervals. This slope, however, becomes $87 \pm 27$ km/nPa when we try to fit all the data points in Figure~\ref{fig:sw}a. Nevertheless, both fittings show a negative correlation between the ionopause altitude and the solar wind dynamic pressure.

Similar ionopause altitude variation with the solar wind dynamic pressure was also observed at Venus by PVO \cite{brace_dynamic_1980}. The trend shown in Figure~\ref{fig:sw}a, however, was not found in \citeA{vogt_ionopause-like_2015}. This can simply be due to the relatively small size of their dataset since this work was done at the very early stage of the MAVEN (Mars Atmosphere and Volatile EvolutioN) mission. Later using MARSIS in situ measurements, \citeA{duru_martian_2020} was actually able to show the similar result as ours that the ionopause altitude decreases while the solar wind dynamic pressure increases. These solar wind effects are also observed in other plasma boundaries at Mars, such as the bow shock, magnetic pileup boundary \cite{edberg_plasma_2009}, ion composition boundary \cite{halekas_structure_2018}, and the photoelectron boundary \cite{withers_morphology_2016, garnier_martian_2017, girazian_effects_2019}.

To test the theory for pressure balance at the ionopause, we plot the total pressure inside the ionopause ($P_\textup{\scriptsize{tot}}=P_\textup{\scriptsize{th}}+P_\textup{\scriptsize{Bi}}$, measured at the ``center point'' of an ionopause as shown in Figure~\ref{fig:alt}) as a function of the normal component of the solar wind dynamic pressure for $\textup{SZA}<65^\circ$ in Figure~\ref{fig:sw}b. We then perform a linear fit (least squares method) for the average pressure in each pressure bin using the formula $P_\textup{\scriptsize{tot}}=aP_{\textup{\scriptsize{sw}} \perp}+b$. Given the large uncertainties and many assumptions we made in calculating $P_\textup{\scriptsize{tot}}$ and $P_\textup{\scriptsize{sw}}$ in section~\ref{sec:pressure}, such as a fixed profile of $T_\textup{\scriptsize{e}}$ and closest $P_\textup{\scriptsize{sw}}$ within 6 hours of ionopause detections, we still find that $a=1.36 \pm 0.44$, a slope somewhat higher than 1 but consistent with unity. The result of the test potentially indicates that the total pressure $P_\textup{\scriptsize{tot}}$ inside the ionopause on average balances the normal component of the solar wind dynamic pressure $P_{\textup{\scriptsize{sw}} \perp}$, in agreement with the similar test based on the simultaneous measurements by ASPERA and MAVEN (Mars Atmosphere and Volatile Evolution) in \citeA{sanchezcano_mars_2020}. Additionally, we find that $b=0.19 \pm 0.06$ nPa, a small offset between $P_\textup{\scriptsize{tot}}$ and $P_{\textup{\scriptsize{sw}} \perp}$, which suggests that $P_\textup{\scriptsize{tot}}$ may slightly exceed $P_{\textup{\scriptsize{sw}} \perp}$ at the location where the ionopause forms, in agreement with the results shown in \citeA{holmberg_maven_2019}.

\subsection{Dependence of Ionopause Thickness on Altitude}
\label{subsec:thickness}

\begin{figure*}
\begin{center}
\includegraphics[width=6.69in]{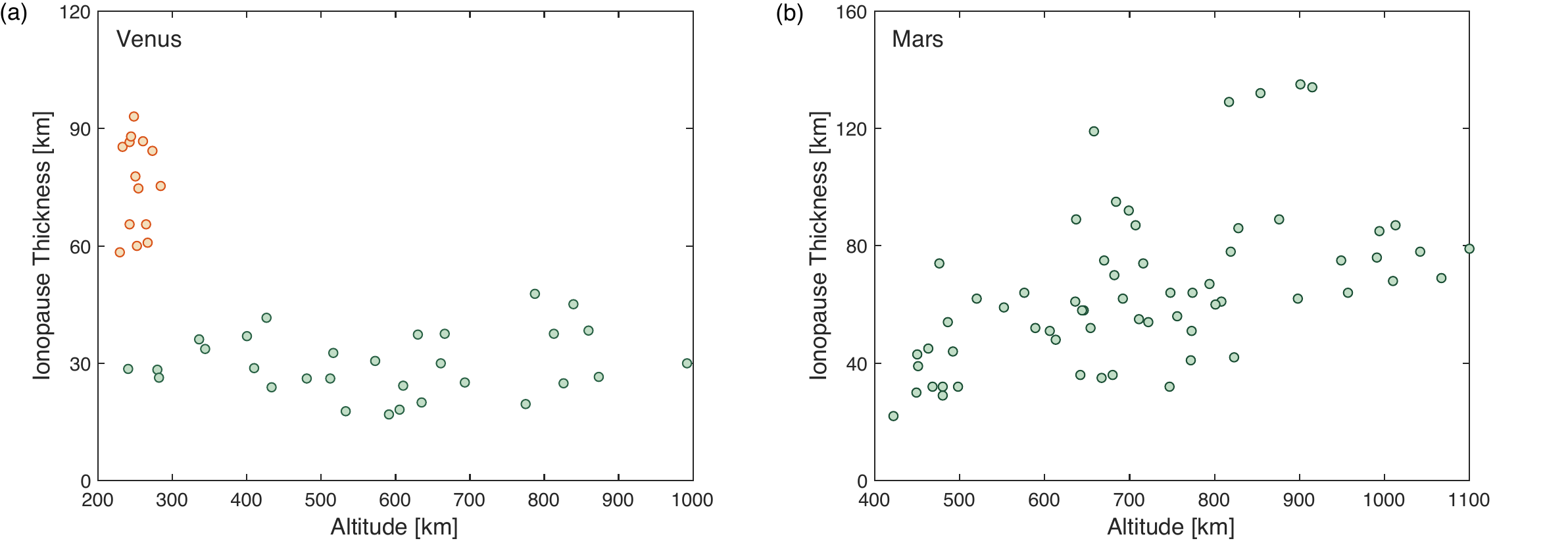}
\caption{(a) Ionopause thickness as a function of altitude at Venus, reproduced from Figure 6 in \citeA{elphic_venus_1981}. The orange and green dots mark the thick ionopauses (thickness $>$ 60 km) forming at low altitudes and thin ionopauses at high altitudes, respectively. (b) Ionopause thickness as a function of altitude at Mars, based on the MARSIS in situ measurements. The ``ionopause altitude'' in (a) and (b) refers to the altitude of the center point of an ionopause, as indicated by the red stars in Figure~\ref{fig:alt}.}
\label{fig:ta}
\end{center}
\end{figure*}

We identify the thickness of a Martian ionopause as the length scale of the steep change in ionospheric plasma density continuously greater than $\Delta n/n> 0.1$. Since the MARSIS in situ technique can only measure the local plasma density once in each frequency sweep period (1.257 s), assuming the local plasma density is $\sim 100 \textup{ cm}^{-3}$ and the vertical speed of MEX is $\sim 3 \textup{ km/s}$, this sorting criteria is equivalent to a minimum density gradient of $\sim 2 \textup{ cm}^{-3}/\textup{km}$. An example illustrating the ionopause thickness is shown in Figure~\ref{fig:alt}. Since the ionopause is essentially a current sheet induced to shield the solar wind magnetic field from penetrating into the upper ionosphere, the thickness of this boundary layer is expected to be on the order of the ion gyroradius scale \cite{cravens_ionopause_1991}. However, there are also other factors that can affect the ionopause thickness, such as diffusion \cite{elphic_venus_1981}.

\begin{figure*}
\begin{center}
\includegraphics[width=6.69in]{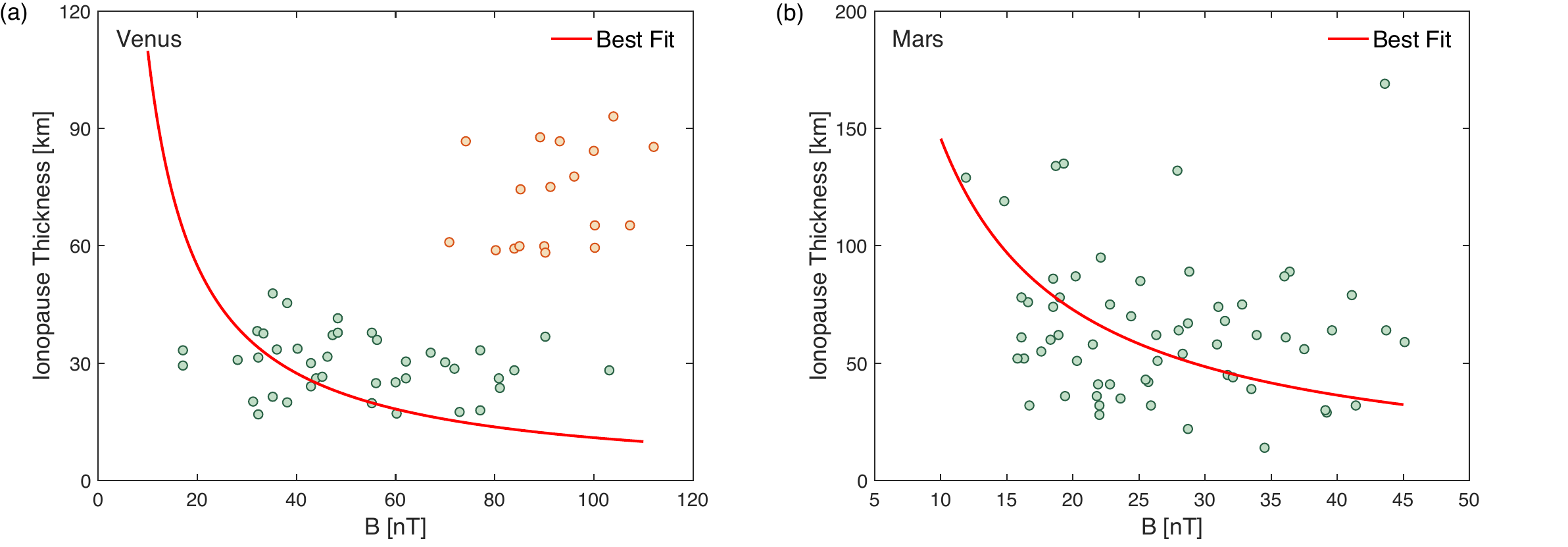}
\caption{(a) Ionopause thickness as a function of magnetic field strength at Venus, reproduced from Figure 5 in \citeA{elphic_venus_1981}. Same as Figure~\ref{fig:ta}a, the orange and green dots represent the ionopauses forming at low and high altitudes, respectively. (b) Ionopause thickness as a function of magnetic field strength at Mars, based on the MARSIS in situ measurements. The red curves in (a) and (b) represent the best fit (least squares method) of the green dots with the formula $d=\gamma \sqrt{m_\textup{\scriptsize{i}}kT_\textup{\scriptsize{i}}}/Be \propto 1/B$. }
\label{fig:tb}
\end{center}
\end{figure*}

At Venus, past observations from PVO showed that the ionopauses can be mainly grouped into two classes based on the mechanisms that determine their thickness -- the thick ionopauses (thickness $>$ 60 km) at low altitudes and thinner ones at high altitudes \cite{elphic_venus_1981}. The PVO measurements of the ionopause thickness as a function of altitude are shown in Figure~\ref{fig:ta}a \cite<reproduced from Figure 6 in>{elphic_venus_1981}. When the ionopause forms at low altitudes (colored in orange), due to relatively high ionospheric density, the ion coulomb collision rate usually dominates the ion gyrofrequency in this boundary layer, making diffusive broadening an important process in determining the ionopause thickness. However, as the altitude increases, the coulomb collision rate falls off rapidly, causing the thickness of the ionopause forming in high altitudes (colored in green) to be simply proportional to the ion gyroradius. Since the variation of the ion temperature is small above 300 km, the thickness of these ionopauses is also inversely proportional to the magnetic field strength \cite{miller_solar_1980}. Figure~\ref{fig:tb}a shows the PVO measurements of the ionopause thickness as a function of the field strength \cite<reproduced from Figure 5 in>{elphic_venus_1981}. If we rewrite the ion gyroradius as $\rho_\textup{\scriptsize{i}}=\sqrt{m_\textup{\scriptsize{i}}kT_\textup{\scriptsize{i}}}/Be$, where $m_\textup{\scriptsize{i}}$ is the ion mass and $e$ is the electron charge, by fitting (least squares method) the thickness of the ionopauses at high altitudes (colored in green) with formula $d=\gamma\rho_\textup{\scriptsize{i}}$ and assuming the temperature of $\textup{O}^+$ ions $T_\textup{\scriptsize{i}} \approx 3000 \pm 500$ K, we find that on average the ionopause thickness is about $\gamma = 5.4 \pm 1.4$ ion gyroradii at Venus \cite{elphic_venus_1981}.

At Mars, however, the ionospheric density is much smaller than that at Venus. Previous studies have shown that the ion coulomb collision rate is only comparable to the ion gyrofrequency in the ionospheric dynamo region that lies between 100 and 250 km in altitude, well below that of the ionopause \cite{withers_theoretical_2008, opgenoorth_day-side_2010}. Therefore, in contrast to the two main mechanisms that affect the thickness of Venusian ionopauses at low and high altitudes, the ion gyroradius scale becomes the most important factor that determines the ionopause thickness at Mars. Figure~\ref{fig:ta}b shows the ionopause thickness as a function of altitude based on the MARSIS in situ measurements. In general, we find that the ionopause thickness increases with altitude at Mars due to lower magnetic field strength at higher altitudes \cite{holmberg_maven_2019}. Another contributing factor to this trend is that higher fields are associated with higher solar wind dynamic pressure, thus thin ionopauses tend to be found in lower altitudes. Figure~\ref{fig:tb}b shows the ionopause thickness as a function of the field strength. If we assume the topside ionospheric composition is about 50\% $\textup{O}^+$ ions and 50\% $\textup{O}_2^+$ ions, and $T_\textup{\scriptsize{i}} \approx  T_\textup{\scriptsize{e}} \approx  3000 \pm 500$ K, by repeating the same procedure as in Figure~\ref{fig:tb}a, we find that the ionopause at Mars has a thickness of $5.8 \pm 1.3$ ion gyroradii, comparable to 5.4 ion gyroradii at Venus \cite{girazian_seasonal_2019}.

\section{Discussion on Ionopause Thickness}
\label{sec:discussion}

\begin{figure*}
\begin{center}
\includegraphics[width=6.69in]{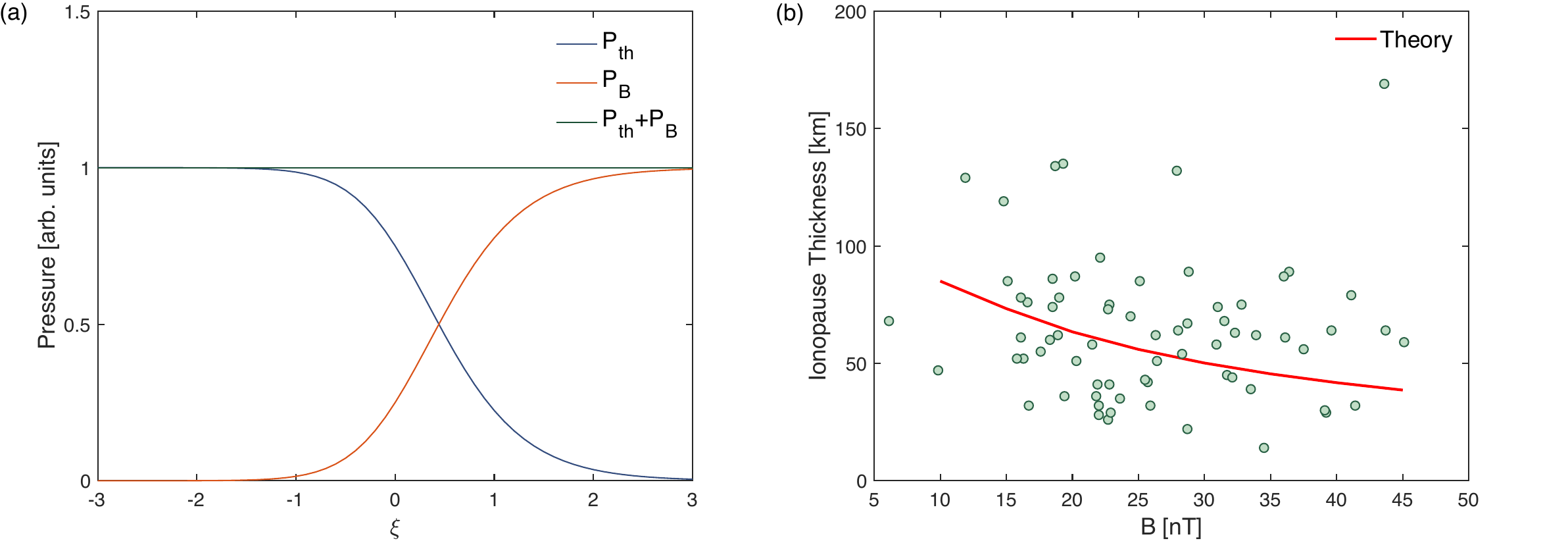}
\caption{(a) Thermal pressure and magnetic pressure profile at the plasma boundary layer based on the magnetic field expression in equation~(\ref{eq:discussion:B}). (b) Theoretical predictions of the ionopause thickness as a function of magnetic field strength (red curve) at Mars, along with the MARSIS in situ observations (green dots). }
\label{fig:discussion}
\end{center}
\end{figure*}


To better understand how the ion gyroradius determines the ionopause thickness at Mars, we can consider the ``thinnest'' boundary layer between a plasma and a vacuum magnetic field \cite{grad_boundary_1961, elphic_venus_1981}. Here we assume that the magnetic field at the ionopause can be approximately represented by the following profile
\begin{equation}
\label{eq:discussion:B}
B=B_0(1+\tanh\xi ),
\end{equation}
where $\xi=(z-z_0)/H$, $z$ is the altitude, $H$ is the arbitrary characteristic length scale of the magnetic field variation, and $B_0$ is the magnetic field strength inside the ionopause ($z=z_0$). Since we are only interested in calculating the ionopause thickness $\Delta z$, we can simply assume $z_0=0$. A steady state plasma boundary requires the total pressure inside the ionopause the same everywhere (Figure~\ref{fig:discussion}a). By solving the pressure balance equation $P_\textup{\scriptsize{th}}(\xi)+P_\textup{\scriptsize{Bi}}(\xi)=P_\textup{\scriptsize{Bi}}(\xi \rightarrow \infty )$ and assuming $T_\textup{\scriptsize{i}} \approx T_\textup{\scriptsize{e}}$, we can obtain the expression for the electron density
\begin{equation}
\label{eq:discussion:ne}
n_\textup{\scriptsize{e}}=\frac{B_0^2}{4\mu _0kT_\textup{\scriptsize{i}}}\left ( 3-\tanh^2\xi-2\tanh\xi \right ).
\end{equation}
The plasma density gradient can thus be computed as
\begin{equation}
\label{eq:discussion:dne}
\frac{dn_\textup{\scriptsize{e}}}{d\xi }=-\frac{B_0^2 }{2\mu _0kT_\textup{\scriptsize{i}}}\textup{ sech}^2\xi (\tanh\xi +1).
\end{equation}
For the thinnest possible boundary profile, \citeA{elphic_venus_1981} found that $H \approx \rho_\textup{\scriptsize{i}}$ when $T_\textup{\scriptsize{i}} \approx T_\textup{\scriptsize{e}}$, $\rho_\textup{\scriptsize{i}}$ being the ion gyroradius. Therefore, by inserting the ionopause identifying criteria $dn_\textup{\scriptsize{e}}/dz > 2 \textup{ cm}^{-3}/\textup{km}$, we are able to calculate the ionopause thickness for various magnetic field strengths as shown in Figure~\ref{fig:discussion}b. Here we adopt the same assumptions made in section~\ref{subsec:thickness}, such as the ionospheric composition at Mars is about 50\% $\textup{O}^+$ ions and 50\% $\textup{O}_2^+$ ions, and $T_\textup{\scriptsize{i}} \approx 3000$ K. It is much to our surprise that these theoretical predictions agree very well with the MARSIS in situ observations, better than the best fit in Figure~\ref{fig:tb}b.

One interesting result from this simple model is that the ratio of the ionopause thickness to ion gyroradius is no longer a constant as we previously assumed in section~\ref{subsec:thickness}; it is now dependent on the magnetic field strength (Table~\ref{table:discussion}). We notice that the ratio ranges from $\gamma=3.4$ at 10 nT to $\gamma=6.8$ at 45 nT, however, on average it is close to our predicted value of 5.8 based on the best fit in Figure~\ref{fig:tb}b. Compared to $\gamma$ being a constant, the field strength dependent profile of $\gamma$ results in a better agreement between the theory and the observations in Figure~\ref{fig:discussion}b, suggesting that the latter might be the actual case scenario that determines the ionopause thickness at Mars. In addition, since this model does not impose constrains on any specific plasma boundary, it can as well be applied to the ionopause at Venus. This explains the coincidence that the ratio of the ionopause thickness to ion gyroradius at Venus is comparable to that at Mars.


\begin{table}
\caption{Ratio of ionopause thickness to ion gyroradius for various magnetic field strengths at Mars: $d=\gamma\rho_\textup{\scriptsize{i}}$, $d$ being the ionopause thickness and $\rho_\textup{\scriptsize{i}}$ the ion gyroradius.}
\label{table:discussion}
\begin{center}
\begin{tabular}{c c c c c c c c c c c}
\hline
$B$ [nT] & 10 & 15 & 20 & 25 & 30 & 35 & 40 & 45 & 50 & 55\\
\hline
$\gamma$  & 3.4  & 4.3 & 5.0 & 5.5 & 5.9 & 6.3 & 6.6 & 6.8 & 7.1 & 7.3\\
\hline
\end{tabular}
\end{center}
\end{table}

\section{Conclusions}
\label{sec:conclusions}

In conclusion, we have investigated the solar wind interaction and pressure balance at the dayside ionopause of Mars using both in situ and remote sounding measurements from the MARSIS instrument. We have found that most of the time the magnetic pressure dominates the thermal pressure to hold off the solar wind at the ionopause at Mars. Only about 13\% of the ionopauses that we examined over a 11-year period are unmagnetized, whereas the unmagnetized ionopauses account for at least 52\% at Venus. Additionally, our analysis has shown that the ionopause altitude decreases as the solar wind dynamic pressure increases at Mars, similar to the altitude variation of the ionopauses at Venus. Finally, we have shown for the first time that the thickness of the ionopauses at Mars is mainly determined by the ion gyromotion, much alike the ionopauses forming in high altitudes at Venus. The ionopauses at Mars are found to have a thickness of about 5.8 ion gyroradii, surprisingly close to the ionopause thickness of 5.4 ion gyroradii at Venus.

\acknowledgments

This work was supported by NASA through Contract No. 1560641 with the Jet Propulsion Laboratory. The MARSIS data used in this study are publicly available through the NASA Planetary Data System (PDS; \url{https://pds-geosciences.wustl.edu/missions/mars_express/marsis.htm}). The PVO OMAG data \cite{kniffin_pvo_1993} and OETP data \cite{theis_pvo_1993} are also publicly available through the NASA PDS. The solar wind moments and derived data products shown in Figures~\ref{fig:sw}, \ref{fig:ta}, and \ref{fig:tb} are archived on \url{zenodo.org} \cite{chu_dayside_2021}.

\bibliography{../refs}

%
%
%
%
%

\end{document}